\documentclass[conference]{IEEEtran}
\IEEEoverridecommandlockouts
\usepackage{amssymb, amsmath, amsthm, amsfonts,graphicx}
\usepackage{algorithm}
\usepackage{algpseudocode}
\usepackage{hyperref}

\newcommand{\commentOut}[1]{}

\ifCLASSINFOpdf
\else
\fi
\begin{document}
%
\title{When Cars Meet Distributed Computing: \\Data Storage
as an Example}

\author{\IEEEauthorblockN{Lewis Tseng\IEEEauthorrefmark{1}\thanks{\IEEEauthorrefmark{1}This work was partially done when Lewis Tseng was with Toyota InfoTechnology Center, USA, Inc.}}
\IEEEauthorblockA{Deaprtment of Computer Science\\Boston College\\Boston, MA, USA\\lewis.tseng@bc.edu}
\and
\IEEEauthorblockN{Takamasa Higuchi}
\IEEEauthorblockA{Toyota InfoTechnology Center,\\ Co., Ltd.\\Tokyo, Japan\\ta-higuchi@jp.toyota-itc.com}
\and
\IEEEauthorblockN{Onur Altintas}
\IEEEauthorblockA{Toyota InfoTechnology Center,\\ USA, Inc.\\Mountain View, CA, USA\\onur@us.toyota-itc.com}
}



%


\maketitle

\begin{abstract}
	\normalsize
As cars are ubiquitous they could play a major role in a next generation communication and computation framework. In the last years, the development of vehicle-to-vehicle communication and vehicle-to-infrastructure communication took huge steps forward and therefore gives us the tools to build ``mobile computing service'' on cars equipped with computation capabilities. Recently, several groups of researchers independently proposed the design of ``vehicular clouds'' that materializes the concept.
In this paper, we introduce a new paradigm of the vehicular clouds, followed by a case study of data storage on top of the proposed cloud.
Finally,  we  present  several  challenges  and  opportunities
in  the  intersection of vehicular clouds and distributed computing.
\end{abstract}


%
\IEEEpeerreviewmaketitle

\section{Introduction}

Cars today are equipped with a rich set of computing, data storage,  communication, and sensor resources in their on-board computer unit.
Based on current standards, not only vehicle-to-infrastructure (V2I) but also Vehicle to Vehicle (V2V) communication are supported. 
Thus, it is believed that cars will play a major role in future Information and Communication Technology systems for supporting applications like smart cities and intelligent traffic systems, e.g., \cite{WiMobCity14,Olariu10,Gerla14}.

Recently, several groups of researchers independently proposed the concept of ``vehicular clouds'' (VC) which brings the mobile cloud model (or edge computing model \cite{Edge16}) to vehicular networks. Eltoweissy et al. were among the first to propose using vehicular clouds (VC) as the backbone of intelligent transportation systems, smart cities, and smart electric power grids \cite{Olariu10}. Lee et al. drafted important design principles of building VC on top of Vehicular Ad-hoc NETwork (VANET) and information-centric networking \cite{Gerla14}. Dressler et al. \cite{WiMobCity14,VNC16_parkedcar,CarSys17_parkedcars} discussed how to provide cloud-like computation and networking service over cars in a parking lot. Independently, Olariu et al. also explored how to distribute MapReduce-like computation onto cars in a parking lot, e.g., \cite{Olariu17_job_completion,airport12}. Onur et al. proposed Car4ICT, a V2V-based mechanism for providing services from individual cars \cite{Car4ICT15}.

Several advantages over the current cloud computing platforms were mentioned: (i) VC allows diverse resources to be pooled and deployed dynamically to serve users with different needs (e.g., on-demand market for computation and communication service from nearby cars), (ii) VC enables autonomy in real-time service sharing and management with lower network latency (if VC is deployed in proximity), and (iii) VC is typically decentralized and peer-to-peer which avoids a single point of failures or market monopoly.

Despite all these efforts, there are still many challenges and open problems.
Prior work has discussed issues ranging from networking \cite{Gerla14}, engineering \cite{survey14} to security \cite{security13}. In this paper, we focus on the issues in the intersection of VC and distributed computing, particularly on issues related to the dynamic and distributed nature of vehicles.  
To address some of the challenges, we introduce a new paradigm of vehicular cloud, namely \textit{Macro-Micro Cloud} (MMC), and later discuss how MMC can be used for data storage service.
We conclude the paper with exciting challenges and opportunities.

\section{Macro-Micro Cloud}



\noindent \textit{Current Limitation}:
The core mechanism of existing vehicular clouds is as follows (e.g., \cite{Car4ICT15,Gerla14}):
(i) A user sends a service request message to a car passing by using V2V or V2I communication; (ii) The request is then forwarded over the communication network to discover a communication path to a vehicle offering the desired service; and (iii) If the user successfully finds a service provider, s/he uses this communication path to exchange service-related data thereafter. Although the mechanism works well for a variety of services as shown in prior works, it is non-trivial to adapt the mechanism for many other services, e.g., long-lasting services tied to a certain geographical area.
Due to vehicle mobility, a user needs to find an alternative service provider whenever the current service provider leaves the neighborhood. This causes huge communication overhead and poor quality of service. Recent work \cite{CarSys17_parkedcars,CarSys17_data} addressed the issue by using a cluster of parked cars to connect to the Internet Clouds. Here, we focus on the design that minimizes the reliance on V2I communication and cars may move around.

\noindent \textit{Intuition of MMC}:
The main goal of Macro-Micro Cloud (MMC) is to reduce the communication complexity and improve quality of service for long-lasting location-based services, since such types of services are important to intelligent traffic systems and smart cities, e.g., computation backbone for Internet-of-Things and dynamic traffic routing. 

Our design is based on the following observations: (i) for location-based services, nearby cars can coordinate with each other to provide the service more efficiently (compared with an opportunistic approach in recent work like \cite{Car4ICT15}), (ii) hierarchical structure allows us to deploy MMC more flexibly and efficiently, and (iii) some hand-off mechanism is necessary for long-lasting services (to handle the case when there is no available cars in the neighborhood).

\noindent \textit{Design of MMC}:
A key idea behind Macro-Micro Cloud (MMC) is the \textit{hierarchical cloud structure}. That is, cars form multiple local clusters, namely micro vehicle clouds, and micro clouds behave as special nodes of the macro vehicle cloud (which is similar to Car4ICT \cite{Car4ICT15}). In micro clouds, cars entering a designated region (e.g., an intersection) dynamically join a micro cloud using a membership management protocol. The micro cloud is logically fixed and \textit{tied to that region}, and the cars belong to the same micro cloud collaborate with other (micro) cloud members to provide services, e.g., continuous monitoring. While individual car may leave and join the designated region in a short period of time, the micro cloud can keep providing its services as other cars or nearby micro clouds in the region can take over the tasks using a suitable hand-off mechanism.
 
Macro cloud is similar to the concept of Car4ICT \cite{Car4ICT15} in which the cloud consists of individual cars that can discover/use the services using V2V communication. In MMC, we use V2V communication to discover and provision services among micro clouds and individual cars and/or users that access the service using other devices like smart phones and smart watches. The following figure illustrates the design: 

\begin{center}
	\includegraphics[scale=0.32]{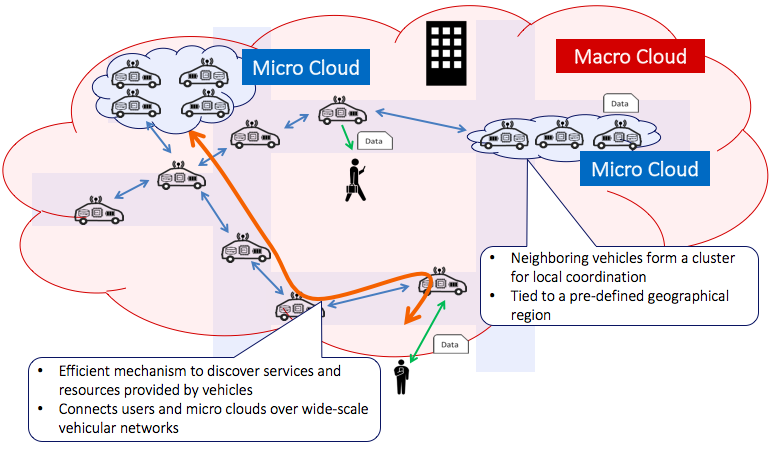}
\end{center}


\noindent \textit{Features of MMC}:

\begin{itemize}
	\item The static nature of micro clouds is especially beneficial to the long-lasting services, as a user does not need to repeatedly send service request messages as in \cite{Car4ICT15,Gerla14}. 
	
	\item Coordination among the cars in micro clouds and the hand-off mechanism among micro clouds supports long-lasting services despite vehicle mobility. 
	
	\item Through macro cloud, MMC provides wide-area service discovery and provisioning.
	
	\item Hierarchical structure makes the deployment flexible. For example, micro clouds at different location may use different hand-off and/or membership management protocols to better accommodate different conditions, e.g., traffic condition, vehicle mobility, services requirements.
\end{itemize}


\noindent \textit{Case Study: Data Storage}

The types of services that can be efficiently provided by MMC is beyond the scope of this paper. Here we discuss one application: how does MMC provide location-based data storage service? Such service could be useful for storing traffic related data or data generated by Internet-of-Things. MMC uses the following steps to provide the storage service:

\begin{itemize}
	\item Identify the locations that are suitable for deploying micro clouds, e.g., a locations that have enough cars passing by.
	\item Using historical data to build a model to estimate the availability of resource (e.g., storage space and bandwidth) for each micro cloud.
	\item Based on the availability model, calculate the amount of data that can be maintained and stored at each micro cloud at a given period of time.
	\item The final step is to implement storage on suitable micro clouds using a mechanism similar to GeoQuorum \cite{GeoQuorum}.\footnote{Instead of quorum-based approach, we used GPS clock to approximate atomicity. A mechanism to ensure atomicity is an interesting open problem.}
\end{itemize}

\section{Challenges and Opportunities}

We categorize the challenges into three groups: 

\begin{itemize}
	\item \textit{Game theoretical analysis}: One of the major issues for vehicular clouds is ``why would cars participate''? 
	Intuitively, vehicular cloud is similar to Peer-to-Peer systems like BitTorrent in which users need to contribute in order to use the services.
	However, such an argument is not sufficient for critical services in intelligent traffic systems, e.g., dynamically routing cars, since in this case, the service is expected to be provided to most cars even if they are not participating. Therefore, we need more rigorous study on the incentive mechanism for forming and participating in vehicular clouds.
	
	\item \textit{Multiplicity of network}: Cars have different communication capabilities, e.g., DSRC, Wi-Fi, LTE, millimeter wave, etc. Each of the communication mechanisms has different characteristics. How to use them efficiently for a certain service is an open problem. For example, using fast and low-bandwidth communication for exchanging frequent overhead messages would be beneficial if the service does not have huge overhead.
	\item \textit{Incentivized mobility}: In many prior works, the algorithm is designed for the worst case, in which cars may move at will. However, it is possible to ``guide'' cars to move in a desirable way to maximize the performance (either using monetary compensation or rewards on using premium services). It is interesting to explore such an incentive in mobility models, and how it affects performance.
\end{itemize}


\bibliographystyle{abbrv}
\bibliography{car} 

\begin{thebibliography}{10}

\bibitem{Car4ICT15}
O.~Altintas, F.~Dressler, F.~Hagenauer, M.~Matsumoto, M.~Sepulcre, and
  C.~Sommer.
\newblock {Making Cars a Main ICT Resource in Smart Cities}.
\newblock In {\em 34th IEEE Conference on Computer Communications (INFOCOM
  2015), International Workshop on Smart Cities and Urban Informatics
  (SmartCity 2015)}, pages 654--659, Hong Kong, China, April 2015. IEEE.

\bibitem{airport12}
S.~Arif, S.~Olariu, J.~Wang, G.~Yan, W.~Yang, and I.~Khalil.
\newblock Datacenter at the airport: Reasoning about time-dependent parking lot
  occupancy.
\newblock {\em IEEE Transactions on Parallel and Distributed Systems},
  23(11):2067--2080, Nov 2012.

\bibitem{WiMobCity14}
F.~Dressler, P.~Handle, and C.~Sommer.
\newblock Towards a vehicular cloud - using parked vehicles as a temporary
  network and storage infrastructure.
\newblock In {\em Proceedings of the 2014 ACM International Workshop on
  Wireless and Mobile Technologies for Smart Cities}, WiMobCity '14, pages
  11--18, New York, NY, USA, 2014. ACM.

\bibitem{Olariu10}
M.~Eltoweissy, S.~Olariu, and M.~Younis.
\newblock {\em Towards Autonomous Vehicular Clouds}, pages 1--16.
\newblock Springer Berlin Heidelberg, Berlin, Heidelberg, 2010.

\bibitem{Olariu17_job_completion}
R.~Florin, P.~Ghazizadeh, A.~G. Zadeh, S.~El-Tawab, and S.~Olariu.
\newblock Reasoning about job completion time in vehicular clouds.
\newblock {\em IEEE Transactions on Intelligent Transportation Systems},
  18(7):1762--1771, July 2017.

\bibitem{VNC16_parkedcar}
F.~Hagenauer, C.~Sommer, T.~Higuchi, O.~Altintas, and F.~Dressler.
\newblock {Using Clusters of Parked Cars as Virtual Vehicular Network
  Infrastructure}.
\newblock In {\em 8th IEEE Vehicular Networking Conference (VNC 2016), Poster
  Session}, pages 126--127, Columbus, OH, December 2016. IEEE.

\bibitem{CarSys17_parkedcars}
F.~Hagenauer, C.~Sommer, T.~Higuchi, O.~Altintas, and F.~Dressler.
\newblock {Parked Cars as Virtual Network Infrastructure: Enabling Stable V2I
  Access for Long-Lasting Data Flows}.
\newblock In {\em 23rd ACM International Conference on Mobile Computing and
  Networking (MobiCom 2017), 2nd ACM International Workshop on Smart,
  Autonomous, and Connected Vehicular Systems and Services (CarSys 2017)},
  Snowbird, UT, October 2017. ACM.
\newblock to appear.

\bibitem{CarSys17_data}
F.~Hagenauer, C.~Sommer, T.~Higuchi, O.~Altintas, and F.~Dressler.
\newblock {Vehicular Micro Clouds as Virtual Edge Servers for Efficient Data
  Collection}.
\newblock In {\em 23rd ACM International Conference on Mobile Computing and
  Networking (MobiCom 2017), 2nd ACM International Workshop on Smart,
  Autonomous, and Connected Vehicular Systems and Services (CarSys 2017)},
  Snowbird, UT, October 2017. ACM.
\newblock to appear.

\bibitem{Gerla14}
E.~Lee, E.~K. Lee, M.~Gerla, and S.~Y. Oh.
\newblock Vehicular cloud networking: architecture and design principles.
\newblock {\em IEEE Communications Magazine}, 52(2):148--155, February 2014.

\bibitem{GeoQuorum}
N.~L. A.~S. S.~Dolev, S.~Gilbert and J.~Welch.
\newblock ``geoquorum: Implementing atomic memory in ad hoc networks''.
\newblock In {\em 17th International Conference on Principles of DIStributed
  Computing}, pages 306--320, 2003.

\bibitem{Edge16}
W.~Shi, J.~Cao, Q.~Zhang, Y.~Li, and L.~Xu.
\newblock Edge computing: Vision and challenges.
\newblock {\em IEEE Internet of Things Journal}, 3(5):637--646, Oct 2016.

\bibitem{survey14}
M.~Whaiduzzaman, M.~Sookhak, A.~Gani, and R.~Buyya.
\newblock A survey on vehicular cloud computing.
\newblock {\em J. Netw. Comput. Appl.}, 40:325--344, Apr. 2014.

\bibitem{security13}
G.~Yan, D.~Wen, S.~Olariu, and M.~C. Weigle.
\newblock Security challenges in vehicular cloud computing.
\newblock {\em IEEE Transactions on Intelligent Transportation Systems},
  14(1):284--294, March 2013.

\end{thebibliography}

\end{document}